\documentclass[11pt,a4paper]{article}
\usepackage{myjheppub}

\begin{document}

\title{First law of black hole mechanics in variable background fields}

\abstract{It is well known that in general theories of gravity with the diffeomorphism
symmetry, the black hole entropy is a Noether charge. But what will happen
if the symmetry is explicitly broken? By investigating the covariant first
law of black hole mechanics with background fields, we show that the Noether
entropy is still applicable due to the local nature of the black hole
entropy. Moreover, motivated by the proposal that the cosmological constant
behaves as a thermodynamic variable, we allow the non-dynamical background
fields to be varied. To illustrate this general formalism, we study
a generic static black brane in the massive gravity.
Using the first law and the scaling argument, we obtain two Smarr formulas.
We show that both of them can be retrieved without relying on the first law,
hence providing a self-consistent check of the theory.}

\author[a,b]{Shao-Feng Wu}
\author[a]{Xian-Hui Ge}
\author[c]{Yu-Xiao Liu}

\affiliation[a]{Department of physics, Shanghai University, Shanghai, 200444, P. R. China}
\affiliation[b]{The Shanghai Key Lab of Astrophysics, Shanghai, 200234, P. R. China}
\affiliation[c]{Institute of Theoretical Physics, Lanzhou University, Lanzhou, 730000, P. R. China}

\emailAdd{sfwu@shu.edu.cn}
\emailAdd{gexh@shu.edu.cn}
\emailAdd{liuyx@lzu.edu.cn}

\maketitle

\flushbottom

\section{Introduction}

When a closed system has a differentiable dynamical symmetry, Noether's
theorem indicates the existence of a corresponding conserved current.
Usually, Noether's theorem is not applicable to the system coupling to the
environment and the extension would be significant as shown in open quantum
systems \cite{Marvian2014}. In field theories, if the environment is inert
or the spontaneous breaking physics appears at a large energy scale, the
system can be mimicked by coupling a background field associated with
certain explicit symmetry breaking.

The gravitational models with explicit diffeomorphism (and Lorentz) breaking
have a long research history. Examples include the massive gravity with a
reference metric \cite{Pauli1939,Veltman1970,deRham2010} and the
Chern-Simons gravity coupling to the axions \cite{Lue1999,Jackiw2003}. Of
particular interest is the recent application of the massive gravity in the
gauge/gravity duality \cite%
{Vegh2013,Davison2013,Blake2013,Baggioli2014,Amoretti2014}, where the
reference metric can imitate the mean-field disorder in realistic materials.

One of the well-known Noether charges in gravitational physics is the
covariant expression of black hole entropy proposed by Wald, with respect to
the diffeomorphism symmetry of the general theories of gravity \cite%
{Wald1,Iyer1994}. The Wald entropy is a \textquotedblleft local,
geometrical\textquotedblright\ quantity on the Killing horizon\footnote{%
By \textquotedblleft local, geometrical\textquotedblright , the entropy is
characterized by a covariant surface term made of the fields appearing in
Lagrangian and their derivatives.}. It is identified from the covariant
first law of black hole mechanics, which is a variational identity built
upon the Hamiltonian that generates the evolution in the phase space of
black hole solutions. Besides its elegant construction and universality, the
success of the Wald entropy is that its higher curvature contribution
precisely agrees with the microscopic entropy computed by state counting in
string or M-theory \cite{Mohaupt2001}.

However, as pointed out by Iyer and Wald in the appendix to Ref. \cite%
{Iyer1994}, the diffeomorphism invariance implies the absence of
\textquotedblleft non-dynamical fields\textquotedblright\ in their
Lagrangian. Nevertheless, they have applied the theories with a
non-dynamical metric (such as the theories of fields in flat spacetimes) to
discuss the canonical energy. But until now, little attention has been paid
on the covariant first law and the Noether entropy with the background
fields. In this paper, we aim to fill this gap. The essential observation is
the following: {the presence of background fields only appends the nonlocal
volume terms to the key variational identity that leads to the first law%
\footnote{%
In the path integral approach to quantum gravity, two configurations related
by a diffeomorphism are physically indistinguishable and should not be
double counted. If any field converts to a background, the two configurations are
different. By contrast, there
is no such qualitative difference which would hinder the construction of the
variational identity with background fields.}, while the black hole entropy
is expected to be defined by local quantities on the horizon\footnote{%
The seeking for local geometrical feature of black hole entropy has
motivated Wald's formalism and was used to present a candidate entropy
definition of dynamical black holes {\cite{Wald1,Iyer1994,Jacobson1993}}.
This feature is inherited from the Bekenstein-Hawking entropy, originally
inspired by the famous teacup gedanken experiment \cite{Bekenstein1972}
which suggests that any black hole horizon should be associated with the
entropy to compensate the hidden information. The black hole entropy
including its thermodynamic and statistical significance can be extended to
more local notion of the causal horizon \cite{Jacobson2003}. Moreover, the
Bekenstein-Hawking entropy enlightened the holographic principle which
states that the information inside a space can be encoded on its boundary.
The celebrated realization of holographic principle by gauge/gravity duality
reassures the local feature of black hole entropy, which is dual to the
dependence of thermal entropy on IR\ physics alone \cite{Liu2014} (We thank
Hong Liu for discussion on this point.). With these in mind, we will use the
local feature to identify the black hole entropy.}}.

The background fields involved in this work are specified as the fields in
the Lagrangian which do not react under the diffeomorphism and whose
equations of motion (EOM) are not imposed. Furthermore, we will allow them
to be varied. In other words, the background fields are able to respond to
the variation of the dynamical fields which are coupled to them. In this
respect, our background fields are more general than the usual
\textquotedblleft prior geometry\textquotedblright\ \cite{Misner}\ or
\textquotedblleft absolute object\textquotedblright\ \cite{Anderson}, which
cannot be changed by changing other fields. The motivation to study the
varied background fields arises from the proposal that the cosmological
constant $\Lambda $ behaves as a thermodynamic variable, the pressure \cite%
{Caldarelli1999,Gibbons2005,WuSQ2006,Urano2009,Kastor2009,Dolan1008}. One
important evidence for this proposal is that the Smarr formula integrated
from the first law with variation $\delta \Lambda $ can be retrieved by the
geometric method \cite{Kastor2009}, hence indicating the existence of the
Killing potential. Recently, in terms of the holographic duality, the origin
of the Smarr formula\ with the pressure for AdS black holes has been
understood as the fact that the free energy of a large $N$ gauge theory only
depends on the color number $N$ via an overall factor $N^{2}$ \cite%
{Karch2015}. Moreover, many interests have been attracted to study the
implication of the extended phase space in black hole phase transitions from
the viewpoint of chemistry \cite{Mann2012,Mann2014,Liu2015,Cao2015}. In our
formalism, one can find that the variable cosmological constant can be
described by the simplest scalar background, which is constant ($\partial
\Lambda =0$) but not fixed ($\delta \Lambda \neq 0$).

As an illustration, we will study a generic static black brane in the
Einstein-Maxwell-Dilaton (EMD) gravity with a reference metric and the
cosmological constant. Both of them will be regarded as varied background
fields. We have interests on the EMD gravity rather than the simple Einstein
gravity since the former is more general: it involves three types of fields
(scalar, vector and tensor) which contribute to the covariant first law with
different forms. Moreover, the massive EMD gravity is very interesting in
recent holographic models since it provides abundant physics in field
theories. For instance, the dilaton is appealing as it features robust
linear in temperature resistivity \cite{Amoretti2014}.

\section{Covariant first law}

Iyer and Wald \cite{Wald1,Iyer1994} have derived the covariant first law of
black hole mechanics by constructing a variational identity. The black hole
entropy is identified with the Noether charge with respect to the
diffeomorphism symmetry. They also pointed out that a number of formulas and
results continue to hold for theories with a non-dynamical metric. However,
they focused on the canonical energy but did not mention what is the black
hole entropy in that case. This can be partially understood since the only
non-dynamical field in their work is the spacetime metric (not the reference
metric) and in the typical non-dynamical spacetime, i.e. the flat spacetime,
the black hole entropy loses physical meaning. In this section, we will not
restrict on a single fixed non-dynamical metric but will extend the
variational identity to involve more general background fields, which can be
scalars, vectors or tensors and all of them are allowed to be varied. At
last, we will identify the black hole entropy from the extended variational
identity in terms of its local nature.

For this purpose, we consider a scalar Lagrangian $L$ as a functional of
some concrete field tensors and their derivatives. These fields are
collectively denoted by $\psi $, including the metric $g^{\mu \nu }$ and
various matter fields (scalar, vector, two-tensor) $\psi =\left( g^{\mu \nu
},\phi ,a_{\mu },b_{\mu \nu }\right) $. Each type of the matter fields can
involve multiple fields, for instance, $\phi =(\phi _{1}$, $\phi _{2}$, $%
\cdots )$, and our formalism below can be generalized directly. The
derivatives of these fields are defined as $\left( R_{\mu \nu \lambda \rho
},\nabla _{\mu }\phi ,\nabla _{\nu }a_{\mu },\nabla _{\lambda }b_{\mu \nu
}\right) $. To track what will be different explicitly if any background
fields are turned on, we will suppose $\psi $ as the dynamical variables at
the beginning, convert one or more variables in $\psi $ to the background
fields in the end, and not impose any EOM in the intermediate steps unless
we state clearly for discussing the on-shell results. Moreover, it should be
mentioned that the different choice of the tensor types (upper or lower
index) of $\psi $ will not change the first law of the
diffeomorphism-invariant theories where all structures are produced
dynamically, but it is not the case when a background field appears. To
accommodate the theory of massive gravity that will be used to exemplify our
general formalism, we adapt the current variables.

We start from the variation of the Lagrangian 4-form%
\begin{equation}
\delta \left( \ast L\right) =\ast E\delta \psi +\mathrm{d}\left( \ast \theta
\right) ,
\end{equation}%
where $\ast $ refers to the Hodge dual and $\theta =\theta \left( \psi
,\delta \psi \right) $ is an one-form. A sum over all variables in $E\delta
\psi $ is understood and the quantity $E$ denotes collectively the EOM $%
E=(E_{\mu \nu }^{(g)},E^{(\phi )},E^{\left( a\right) \mu },E^{(b)\mu \nu })$
with respect to $\psi $. When the theory is on shell, $E=0$ for any
dynamical variables. In the Appendix, we will list the explicit expressions
of the EOM and some lengthy tensors appeared below which can be derived
paralleling Ref. \cite{Maeda1998} but with different variables.

In Ref. \cite{Wald1,Iyer1994}, the Lagrangian is assumed to be
diffeomorphism invariant, that is%
\begin{equation}
\mathrm{L}(f^{\ast }\psi )=f^{\ast }\mathrm{L}\left( \psi \right)
\label{diff0}
\end{equation}%
where $\mathrm{L=}\ast L$ and $f$ denotes any diffeomorphism map. Since the
pullback $f^{\ast }$ does not act on the background fields, there is no
dependence of background fields in the Lagrangian. Equation (\ref{diff0})
implies%
\begin{equation}
\delta _{\xi }\left( \ast L\right) =\mathrm{d}i_{\xi }\left( \ast L\right) =%
\pounds _{\xi }\left( \ast L\right) ,  \label{diff1}
\end{equation}%
where $\delta _{\xi }$ is the variation induced by the diffeomorphism along
any vector $\xi $, whilst $i_{\xi }$ and $\pounds _{\xi }$ denote the
relevant contraction and Lie derivative. In this work, we would like to
figure out the breaking of the diffeomorphism symmetry. So we represent eq. (%
\ref{diff1}) as an equivalent form%
\begin{equation}
0=\pounds _{\xi }L-\xi ^{\mu }\nabla _{\mu }L=P_{\mu \nu }\nabla ^{\nu }\xi
^{\mu }  \label{diff2}
\end{equation}%
but keep the tensor $P_{\mu \nu }$ nonvanishing at this time. Note that $%
P_{\mu \nu }$ is complicated, see eq. (\ref{puv}). Furthermore, one needs to
define a current
\begin{equation}
j_{\xi }=\ast \theta \left( \psi ,\pounds _{\xi }\psi \right) -i_{\xi
}\left( \ast L\right)
\end{equation}%
which satisfies%
\begin{equation}
\mathrm{d}j_{\xi }=-\ast E\pounds _{\xi }\psi .
\end{equation}%
On shell, this current is conserved and has been called as the Noether
current associated with the diffeomorphism symmetry in the sense of \cite%
{Wald1990}. One can prove%
\begin{equation}
\theta ^{\beta }\left( \psi ,\pounds _{\xi }\psi \right) -\xi ^{\beta
}L=\nabla _{\alpha }Q_{\xi }^{\beta \alpha }+\xi ^{\mu }\left( \tilde{E}%
_{\mu }^{\;\beta }+P_{\mu }^{\;\beta }\right) ,
\end{equation}%
where the called Noether potential is given by%
\begin{equation}
Q_{\xi }^{\beta \alpha }=2\left( X^{\alpha \beta \mu \nu }\nabla _{\mu }\xi
_{\nu }-2\xi _{\nu }\nabla _{\mu }X^{\alpha \beta \mu \nu }+\xi ^{\nu }%
\tilde{Q}_{\nu }^{\;\beta \alpha }\right) .  \label{Q}
\end{equation}%
Here we have defined a four-tensor by the derivative $X^{\alpha \beta \mu
\nu }=\partial L/\partial R_{\alpha \beta \mu \nu }$, a three-tensor $\tilde{%
Q}_{\nu }^{\;\beta \alpha }$ composed of the fields $\left( a_{\mu },b_{\mu
\nu }\right) $ and the derivatives $(\partial L/\partial \nabla _{\nu
}a_{\mu },\partial L/\partial \nabla _{\lambda }b_{\mu \nu })$, and a
two-tensor made by:%
\begin{equation}
\tilde{E}_{\mu \beta }=2E_{\mu \beta }^{(g)}-E_{\beta }^{\left( a\right)
}a_{\mu }-E_{\alpha \beta }^{(b)}b_{\;\mu }^{\alpha }-E_{\beta \alpha
}^{(b)}b_{\mu }^{\;\alpha }.
\end{equation}%
The one-form $\theta \left( \psi ,\delta \psi \right) $ can induce a general
symplectic form%
\begin{equation}
\Omega \left( \psi ,\delta _{1}\psi ,\delta _{2}\psi \right) =\delta _{1}
\left[ \ast \theta \left( \psi ,\delta _{2}\psi \right) \right] -\delta _{2}%
\left[ \ast \theta \left( \psi ,\delta _{1}\psi \right) \right] ,
\end{equation}%
where the two variations are not specified. In the phase space stretched by
the solutions to the EOM, the variation of the Hamiltonian generating the
flow along $\xi $ is related to the special symplectic form%
\begin{equation}
\delta H_{_{\xi }}=\int_{\Sigma }\Omega \left( \psi ,\delta \psi ,\pounds %
_{\xi }\psi \right) ,
\end{equation}%
where the integral is carried on a Cauchy surface $\Sigma $, which connects
the horizon cross section with the spatial infinity. This symplectic form
can be recast as%
\begin{eqnarray}
\Omega \left( \psi ,\delta \psi ,\pounds _{\xi }\psi \right) \!\! &=&\!\!%
\mathrm{d}\big[\delta \left( \ast Q_{\xi }\right) -i_{\xi }\left( \ast
\theta \right) \big]  \nonumber \\
&&+\delta \big(\ast i_{\xi }\tilde{E}\big)+i_{\xi }\big(\ast E\delta \psi %
\big)+\delta \big(\ast i_{\xi }P\big).  \label{Omega}
\end{eqnarray}%
Suppose that $\xi $ is the Killing vector which generates a symmetry
inducing $\pounds _{\xi }\psi =0$. Then one immediately has $\delta H_{_{\xi
}}=0$, which yields a variational identity%
\begin{equation}
\int_{\partial \Sigma }\!\!\delta \left( \ast Q_{\xi }\right) \!-i_{\xi
}\left( \ast \theta \right) \!+\!\int_{\Sigma }\!\!\delta \left( \ast i_{\xi
}\tilde{E}\right) \!+\!i_{\xi }\left( \ast E\delta \psi \right) +\delta
\left( \ast i_{\xi }P\right) =0.  \label{dH1}
\end{equation}%
Equation (\ref{dH1}) is the essential result of this section. Some remarks
are in order.

(i) One can recover the covariant first law of diffeomorphism-invariant
theories based on eq. (\ref{dH1}), as it should be. In this case, by
imposing all the EOM and $P_{\mu \nu }=0$, eq. (\ref{dH1}) is reduced to%
\begin{equation}
\int_{\partial \Sigma }\delta \left( \ast Q_{\xi }\right) -i_{\xi }\left(
\ast \theta \right) =0.  \label{surface1}
\end{equation}%
Consider a stationary black hole with the Killing vector $\xi =\mathrm{t}%
+\Omega _{H}\varphi $, where $\mathrm{t}$ denotes the time translation, $%
\Omega _{H}$ the angular velocity and $\varphi $ the angular rotation.
Equation (\ref{surface1}) can be written as the first law \cite%
{Wald1,Iyer1994}%
\begin{equation}
T\delta S=\delta \mathcal{E}+\Omega _{H}\delta \mathcal{J}.  \label{1st}
\end{equation}%
Here $T$ is the Hawking temperature. The black hole entropy is nothing but
the Noether charge, defined by local geometrical quantities,%
\begin{equation}
S=2\pi \int_{B}\left. \ast Q_{\xi }\right\vert _{\xi \rightarrow 0,\;\nabla
_{\mu }\xi _{\nu }\rightarrow n_{\mu \nu }},  \label{S}
\end{equation}%
where $B$ denotes the bifurcation horizon, $n_{\mu \nu }$ is its binormal,
and any reference to the Killing vector (that is nonlocal) was eliminated. $%
\delta \mathcal{E}$ and $\delta \mathcal{J}$ denote the variations of energy
and angular momentum, respectively\footnote{$\delta \mathcal{E}$ and $\delta
\mathcal{J}$ are not integrable in general unless specific boundary
conditions of field variables are imposed. Various black hole hairs are
possibly involved in $\delta \mathcal{E}$.}.

(ii) In the appendix to Ref. \cite{Iyer1994}, Iyer and Wald studied the
theories in a non-dynamical spacetime. The spacetime metric $g^{\mu \nu }$
is the only background field, and it is fixed. From eq. (\ref{dH1}) and eq. (%
\ref{S}), one can see that the variation of the would-be entropy of black
holes is vanishing in that case. Thus the first law cannot be well defined.
In the rest of our work, we will regard the spacetime metric as a dynamical
variable. Our theory still breaks the diffeomorphism symmetry if some matter
fields are the background fields.

(iii) One may notice that a great simplification from eq. (\ref{dH1}) to eq.
(\ref{surface1}) is the vanishing of the complicated tensor $P_{\mu \nu }$,
which has been ascribed to the diffeomorphism symmetry of the Lagrangian, as
Ref. \cite{Wald1,Iyer1994} did. However, we would like to stress that one
can keep eq. (\ref{diff2}) holding if the Lagrangian is a scalar under the
general coordinate transformation. On the contrary, the diffeomorphism
symmetry, which not only conveys the information of general coordinate
invariance but also implies that the theory is free of \textquotedblleft
prior geometry\textquotedblright , is sufficient but not necessary for that%
\footnote{%
In Appendix B of the textbook \cite{Carroll}, one can find a wonderful
discussion on the difference between the diffeomorphism invariance and the
general coordinate invariance.}. As a result, we still have $P_{\mu \nu }=0$
regardless the presence of background fields\footnote{%
Given a concrete Lagrangian that is a scalar and involves some background
fields, one can check $P_{\mu \nu }=0$ using the EOM. The difference between
$P_{\mu \nu }=0$ and $E_{\mu \nu }^{(g)}\!\!=0$ might be interesting for
some readers.}. Moreover, all the stuff below that will be derived based on $%
P_{\mu \nu }=0$ can be dubbed as the one with respect to general coordinate
invariance, instead of the diffeomorphism symmetry.

(iv) Equation (\ref{dH1}) exhibits that the background fields and their
variations would add two volume integrals on the Cauchy surface $\Sigma $ in
the variational identity. In particular, the first volume integral may be
nonvanishing even when all background fields are fixed. And the second one
will be different for various tensor types. If one turns off all the
background fields except a special scalar (i.e. the cosmological constant)
that is allowed to be varied, eq. (\ref{dH1}) will be reduced to the
variational identity constructed in \cite{Urano2009}.

Using eq. (\ref{dH1}), one can rewrite the first law (\ref{1st}) with the
different $\delta \mathcal{E}$ and $\delta \mathcal{J}$:%
\begin{eqnarray*}
\delta \mathcal{E}\!\! &=&\!\!\int_{\infty }\!\!\delta \big(\ast Q_{\mathrm{t%
}}\big)\!-\!i_{\mathrm{t}}\big(\ast \theta \big)\!+\!\int_{\Sigma }\delta %
\big(\ast i_{\mathrm{t}}\tilde{E}\big)\!+\!i_{\mathrm{t}}\big(\ast E\delta
\psi \big), \\
\delta \mathcal{J}\!\! &=&\!\!\int_{\infty }\!\!\delta \big(\ast Q_{\varphi }%
\big)\!-\!i_{\varphi }\big(\ast \theta \big)\!+\!\int_{\Sigma }\delta \big(%
\ast i_{\varphi }\tilde{E}\big)\!+\!i_{\varphi }\big(\ast E\delta \psi \big).
\end{eqnarray*}%
The presence of the background fields does not change the Wald entropy
simply because the black hole entropy is expected to be localized on the
horizon but the corrections in eq. (\ref{dH1}) are volume terms. The volume
terms are naturally attributed to $\delta \mathcal{E}$ and $\delta \mathcal{J%
}$, which is reminiscent of the known \textquotedblleft physical
process\textquotedblright\ version of the first law \cite{Gao2001}.
Actually, the \textquotedblleft key ingredient\textquotedblright\ in that
analysis, i.e. a general formula (eqs. (4) and (5) in \cite{Gao2001}) for
the variation of the mass and the angular momentum, can be deduced from eq. (%
\ref{dH1}) by some operations: (i) turn off all background fields, (ii) turn
on the energy-momentum source $\delta T_{\mu }^{\beta }$ and charge-current
source $\delta j^{\beta }$, (iii) suppose the variation $\delta $ in eq. (%
\ref{dH1}) as the linear perturbation caused by the sources, and (iv)
consider $\Sigma $ as the unperturbed spacetime. Then one can obtain%
\begin{equation}
E\delta \psi \!\!=\!\!E_{\text{unperturbed}}\left( \psi _{\text{perturbed}%
}\!-\!\psi _{\text{unperturbed}}\right) \!\!=\!\!0,  \label{phyfirst1}
\end{equation}%
\begin{equation}
\delta \tilde{E}\!\!_{\mu \beta }\!\!=\!\!\!\!\left[ 2E_{\mu \beta
}^{(g)}\!-\!E_{\beta }^{\left( a\right) }a_{\mu }\right] _{\text{perturbed}%
}\!\!=\!\!\delta T_{\mu \beta }\!+\!a_{\mu }\delta j_{\beta },
\label{phyfirst2}
\end{equation}%
where we have used\ the EOM with the sources%
\[
2E_{\mu \beta }^{(g)}=\delta T_{\mu \beta },\;E_{\beta }^{\left( a\right)
}=-\delta j_{\beta }.
\]%
Equation (\ref{dH1}) with eqs. (\ref{phyfirst1}) and (\ref{phyfirst2}) can
recover eq. (33) in \cite{Gao2001}, which further leads to the mentioned
general formula.

\section{Massive gravity}

In this section, we will illustrate the covariant first law in massive EMD
gravity. There are five fields, including the spacetime metric $g^{\mu \nu }$%
, the reference metric $b_{\mu \nu }$, the gauge potential $a_{\mu }$, the
dilaton field $\phi $, and the cosmological constant $\Lambda $. We will
take $b_{\mu \nu }$ and $\Lambda $ as two background fields in which one is
a two-tensor and the other is a special scalar. We will assume that both of
them can be varied.

Consider the gravity theory described by the EMD Lagrangian%
\begin{equation}
L_{0}=R-2\Lambda -\frac{1}{2}\left( \partial \phi \right) ^{2}-\frac{1}{4}%
Z(\phi )F^{2}-V(\phi )  \label{L}
\end{equation}%
plus a graviton mass term $L_{1}=U(b_{\mu \nu }g^{\mu \nu })$ \cite%
{Baggioli2014}. Here $R$ is the Ricci scalar and $F$ is the Maxwell field.
The function form of the scalar potential $V$, the effective electromagnetic
coupling $Z$, and the potential $U$ for the reference metric will be
specified latter. Assumed to be projected only on the spatial coordinates $%
x^{i}$, the reference metric is given by $b_{\mu \nu }=c^{2}\delta _{\mu
}^{i}\delta _{\nu }^{j}\delta _{ij}$, where $c$ is a parameter. Note we have
set $16\pi G=1$ for brevity. We will study a generic static black brane with
the metric%
\begin{equation}
ds^{2}=-h(r)dt^{2}+\frac{1}{f(r)}dr^{2}+r^{2}\left( dx^{2}+dy^{2}\right) ,
\label{ansatz}
\end{equation}%
and the gauge potential $A=a_{t}(r)dt.$ The independent EOM can be written as%
\begin{eqnarray}
\frac{c^{2}h}{r^{2}}-\frac{fh}{r^{2}}+\frac{1}{2}hV-\frac{3}{4}\frac{%
Q_{e}^{2}h}{r^{4}Z}+\frac{fh^{\prime }}{r}+\frac{1}{2}f^{\prime }h^{\prime }-%
\frac{fh^{\prime 2}}{2h}+\frac{1}{4}fh\phi ^{\prime 2}+fh^{\prime \prime }
&=&0,  \nonumber \\
-c^{2}+f+\frac{1}{2}r^{2}V+\frac{Q_{e}^{2}}{4r^{2}Z}+rf^{\prime }+\frac{1}{4}%
r^{2}f\phi ^{\prime 2} &=&0,  \nonumber \\
-V^{\prime }+\frac{Q_{e}^{2}Z^{\prime }}{2r^{4}Z^{2}}+\frac{2f\phi ^{\prime
2}}{r}+\frac{1}{2}f^{\prime }\phi ^{\prime 2}+\frac{fh^{\prime }\phi
^{\prime 2}}{2h}+f\phi ^{\prime }\phi ^{\prime \prime } &=&0,  \nonumber \\
-c^{2}+f+\frac{1}{2}r^{2}V+\frac{Q_{e}^{2}}{4r^{2}Z}+\frac{1}{2}rf^{\prime }+%
\frac{rfh^{\prime }}{2h} &=&0,
\end{eqnarray}%
where the prime denotes the derivative with respect to $r$ and the electric
charge $Q_{e}=Zr^{2}\sqrt{f/h}a_{t}^{\prime }$. The extended variational
identity (\ref{dH1}) includes the usual surface terms%
\begin{equation}
\!\!\delta \big(\ast Q_{_{\xi }}\big)\!-\!i_{_{\xi }}\big(\ast \theta \big)%
=\!-2r\sqrt{\frac{h}{f}}\delta f-r^{2}\sqrt{hf}\phi ^{\prime }\delta \phi
-a_{t}\delta Q_{e}\!  \label{CCEMD}
\end{equation}%
and the new volume terms%
\begin{eqnarray}
\big[i_{_{\xi }}\tilde{E}\big]_{\beta } &=&-E_{\alpha \beta
}^{(b)}b_{\;t}^{\alpha }-E_{\beta \alpha }^{(b)}b_{t}^{\;\alpha }=0,
\nonumber \\
\big[i_{_{\xi }}\left( \ast E\delta \psi \right) \big]_{\nu \lambda \rho }
&=&\varepsilon _{t\nu \lambda \rho }\Big[2U^{\prime }r^{-2}\delta c^{2}-\Big(%
2+\frac{\partial V}{\partial \Lambda }\Big)\delta \Lambda \Big].
\end{eqnarray}%
As a result, we have a conserved quantity independent with $r$:%
\begin{equation}
\delta H=\!-2r\sqrt{\frac{h}{f}}\delta f\!-\!r^{2}\sqrt{hf}\phi ^{\prime
}\delta \phi \!-\!a_{t}\delta Q_{e}\!+\!\int \left[ 2U^{\prime }\delta
c^{2}\!-\!r^{2}\left( 2+\frac{\partial V}{\partial \Lambda }\right) \delta
\Lambda \right] \sqrt{\frac{h}{f}}dr\!.
\end{equation}

Although not necessarily, the potential of the reference metric in
holographic models is usually assumed to be $L_{1}=\left( \mathrm{Tr}%
K\right) ^{2}-\mathrm{Tr}K^{2}$ \cite{Vegh2013}, where the matrix $K$ is
defined by a matrix square root $K_{\;\nu }^{\mu }=\sqrt{g^{\mu \lambda
}b_{\lambda \nu }}$, following the same form as in the standard dRGT massive
gravity \cite{deRham2010}. In the following, we will use this special
potential, which is equivalent to set $U^{\prime }=1$ in the current
situation. The more general potential will not change our results
qualitatively. To go ahead, we need to specify the behavior of the dynamical
fields near horizon and boundary. This is enough to derive the first law by
applying the variational identity, even though the explicit solutions are
not known. Similar process can be found, for instance, in \cite{Lu2014}. The
solutions near the horizon can be given by%
\begin{equation}
f(r)=f_{1}(r-r_{0})+\cdots ,\;h(r)=h_{1}(r-r_{0})+\cdots ,\;\phi (r)=\phi
_{0}+\cdots ,
\end{equation}%
where $r_{0}$ denotes the horizon location. From the EOM, the solutions at
the boundary can be solved with the form\footnote{%
Here we focus on the \textquotedblleft standard\textquotedblright\ case with
$0<\sigma <1$, and assume $Z(\phi )=1+Z_{1}\phi ^{2}+\cdots $, $V(\phi )=%
\frac{1}{2}m^{2}\phi ^{2}+\gamma _{4}\phi ^{4}+\cdots $ for simplicity. Note
that the negative $m^{2}$ is allowed provided that it does not violate the
Breitenlohner-Freedman bound \cite{Breitenlohner1982}.}:
\begin{eqnarray}
f(r) &=&\frac{r^{2}}{l^{2}}+c^{2}+\frac{f_{1}}{r^{1-\sigma }}-\frac{f_{2}}{r}%
+\frac{f_{3}}{r^{1+\sigma }}+\frac{Q_{e}^{2}}{4r^{2}}+\cdots ,  \nonumber \\
h(r) &=&\frac{r^{2}}{l^{2}}+c^{2}-\frac{\mu }{r}+\frac{Q_{e}^{2}}{4r^{2}}%
+\cdots ,  \nonumber \\
\phi (r) &=&\frac{\phi _{1}}{r^{\left( 3-\sigma \right) /2}}+\frac{\phi _{2}%
}{r^{\left( 3+\sigma \right) /2}}+\cdots ,
\end{eqnarray}%
where $\mu $ and $\phi _{1,2}$ are some free parameters and%
\begin{eqnarray}
\sigma &=&\sqrt{4m^{2}l^{2}+9},\;\;\;\;\;\;\;\;\;\;f_{1}=\left( 3-\sigma
\right) \phi _{1}^{2}/(8l^{2}),  \nonumber \\
f_{2} &=&\mu -\frac{\left( 9-\sigma ^{2}\right) \phi _{1}\phi _{2}}{12l^{2}}%
,\;f_{3}=\left( 3+\sigma \right) \phi _{1}^{2}/(8l^{2}).
\end{eqnarray}%
Close to the horizon and the boundary, the variational identity can be
expanded. At leading order, they are
\begin{eqnarray}
\left. \delta H\right\vert _{\mathrm{horizon}} &=&T\delta S+\delta \Lambda
\Phi _{\Lambda },  \nonumber \\
\left. \delta H\right\vert _{\mathrm{boundary}} &=&\delta M-\Phi _{e}\delta
Q_{e}+\Phi _{\phi }\delta \Lambda -\frac{\sigma }{96\pi l^{2}}\big[(3-\sigma
)\phi _{1}\delta \phi _{2}-(3+\sigma )\phi _{2}\delta \phi _{1}\big],
\end{eqnarray}%
where we have defined the gravitational mass (density), entropy, temperature
\begin{equation}
M=2\mu ,\;S=4\pi r_{0}^{2},\;T=\frac{1}{4\pi }\sqrt{f^{\prime
}(r_{0})h^{\prime }(r_{0})},
\end{equation}%
and two local potentials as well as two nonlocal potentials%
\begin{eqnarray}
\Phi _{e}\!\! &=&\!\!a_{t}(\infty ),~~~~~~~\ \Phi _{\phi }=\frac{9-\sigma
^{2}}{576\pi }\phi _{1}\phi _{2},  \nonumber \\
\Phi _{c}\!\! &=&\!\!\int_{r_{0}}2\sqrt{\frac{h}{f}}dr,~\;\Phi _{\Lambda
}=\int_{r_{0}}\left( 2+\frac{\partial V}{\partial \Lambda }\right) \sqrt{%
\frac{h}{f}}r^{2}dr.
\end{eqnarray}%
Here \textquotedblleft $\int_{r_{0}}$\textquotedblright\ means to drop any
terms relevant to the integral upper limit $\infty $. These terms are absent
in the first law because they are divergent and exactly cancel other
divergent terms from eq. (\ref{CCEMD}). Note $\partial V/\partial \Lambda
=V/\Lambda \neq 0$ since we have fixed the expansion coefficients in $%
V\left( \phi \right) $ multiplying $l^{2}$ (like $m^{2}l^{2}$) as
dimensionless constants. Then the first law can be obtained by matching the
variational identity near the horizon and at boundary%
\begin{eqnarray}
T\delta S &\mathcal{=}&\delta M-\Phi _{e}\delta Q_{e}+\left( \Phi _{\phi
}-\Phi _{\Lambda }\right) \delta \Lambda +\Phi _{c}\delta c^{2}  \nonumber \\
&&-\frac{\sigma }{96\pi l^{2}}\big[(3-\sigma )\phi _{1}\delta \phi
_{2}-(3+\sigma )\phi _{2}\delta \phi _{1}\big].  \label{first law}
\end{eqnarray}

The first law (\ref{first law}) implies some interesting relations among the
variables. In terms of the dimensional analysis, it is easy to see that the
black brane solution is scale invariant, i.e. the field configurations are
homogeneous functions (with order zero), if the radial coordinate and the
parameters transform as\footnote{%
Note that we are studying the static and isotropic solution so the scaling
of coordinates $t$, $x$, and $y$ are not relevant.}%
\begin{eqnarray}
r &\rightarrow &\lambda r,\;\mu \rightarrow \lambda \mu ,\;Q_{e}\rightarrow
\lambda Q_{e},  \nonumber \\
\;l &\rightarrow &\lambda l,\;\phi _{1}\rightarrow \lambda ^{\frac{3-\sigma
}{2}}\phi _{1},\;\phi _{2}\rightarrow \lambda ^{\frac{3+\sigma }{2}}\phi
_{2}.
\end{eqnarray}%
In terms of the scale invariance and eq. (\ref{first law}), one can take the
gravitational mass as a homogeneous function
\begin{equation}
M(\lambda ^{2}S,\lambda Q_{e},\lambda ^{-2}\Lambda ,\lambda
^{0}c^{2},\lambda ^{\frac{3-\sigma }{2}}\phi _{1},\lambda ^{\frac{3+\sigma }{%
2}}\phi _{2})=\lambda M(S,Q_{e},\Lambda ,c^{2},\phi _{1},\phi _{2}).
\label{M1}
\end{equation}%
Acting the derivative $\partial _{\lambda }$ and setting $\lambda =1$ at
last gives%
\begin{equation}
2S\frac{\partial M}{\partial S}+Q_{e}\frac{\partial M}{\partial Q_{e}}%
-2\Lambda \frac{\partial M}{\partial \Lambda }+\frac{3-\sigma }{2}\phi _{1}%
\frac{\partial M}{\partial \phi _{1}}+\frac{3+\sigma }{2}\phi _{2}\frac{%
\partial M}{\partial \phi _{2}}=M.
\end{equation}%
Using eq. (\ref{first law}) again, we obtain a Smarr formula%
\begin{equation}
M=2TS+Q_{e}\Phi _{e}+2\Lambda \left( \Phi _{\phi }-\Phi _{\Lambda }\right) .
\label{Smarr1}
\end{equation}%
Note that since $c^{2}$ does not transform under the rescaling, eq. (\ref%
{Smarr1}) remains the same if one sets $\delta c^{2}=0$ at the beginning.
Interestingly, there is another scale invariance for which $\Lambda $ is not
rescaled:%
\begin{eqnarray}
r &\rightarrow &\lambda r,\;\mu \rightarrow \lambda ^{3}\mu
,\;Q_{e}\rightarrow \lambda ^{2}Q_{e},  \nonumber \\
c &\rightarrow &\lambda c,\;\phi _{1}\rightarrow \lambda ^{\frac{3-\sigma }{2%
}}\phi _{1},\;\phi _{2}\rightarrow \lambda ^{\frac{3+\sigma }{2}}\phi _{2}.
\end{eqnarray}%
Under this scaling transformation, the field configurations are homogeneous
functions and the EOM are not changed. Compared with eq. (\ref{M1}), now the
gravitational mass behaves as a different homogeneous function%
\begin{equation}
M(\lambda ^{2}S,\lambda ^{2}Q_{e},\lambda ^{0}\Lambda ,\lambda
^{2}c^{2},\lambda ^{\frac{3-\sigma }{2}}\phi _{1},\lambda ^{\frac{3+\sigma }{%
2}}\phi _{2})=\lambda ^{3}M(S,Q_{e},\Lambda ,c^{2},\phi _{1},\phi _{2}),
\end{equation}%
which yields a different Smarr formula%
\begin{equation}
3M=2\left( TS+Q_{e}\Phi _{e}-c^{2}\Phi _{c}\right) .  \label{Smarr2}
\end{equation}%
Similarly, one can set $\delta \Lambda =0$ at the beginning which will not
change eq. (\ref{Smarr2}).

In the previous derivation of eq. (\ref{Smarr1}) and eq. (\ref{Smarr2}),
either $\delta \Lambda $ or $\delta c^{2}$ has to be nonvanishing in the
first law. As a self-consistent check, we will prove both of the Smarr
formulas without using the first law. Before doing this, we note that one
can use the explicit solution found in a special case \cite{Cai2014} to
check eqs. (\ref{Smarr1}) and (\ref{Smarr2}).

Now consider the Einstein equation and the Killing equation which can lead to%
\begin{equation}
\nabla _{\mu }\left[ \left( T^{\mu \nu }-\frac{1}{2}g^{\mu \nu }T\right) \xi
_{\nu }\right] =\nabla _{\mu }\left( 2R^{\mu \nu }\xi _{\nu }\right) =0.
\end{equation}%
Following the geometric method to derive the Smarr formula \cite{Kastor2009}%
, where the key ingredient is the construction of the Killing potential, one
can build up a new conserved tensor, at least locally:%
\begin{equation}
Q_{1}^{\mu \nu }=\nabla ^{\mu }\xi ^{\nu }-\Phi _{1}^{\mu \nu }.
\end{equation}%
Here we have defined the generalized Killing potential which is determined by%
\begin{equation}
2\nabla _{\nu }\Phi _{1}^{\mu \nu }=\left( T^{\mu \nu }-\frac{1}{2}g^{\mu
\nu }T\right) \xi _{\nu }.
\end{equation}%
For the massive gravity, $Q_{1}$ has the nonvanishing components
\begin{equation}
Q_{1}^{tr}=-Q_{1}^{rt}=-\frac{fh^{\prime }}{2h}+\frac{1}{4r^{2}}\sqrt{\frac{f%
}{h}}\left( Q_{e}a_{t}-2\Lambda \tilde{\Phi}_{\Lambda }\right) ,
\end{equation}%
where $\tilde{\Phi}_{\Lambda }=\int \left( 2+V/\Lambda \right) \sqrt{h/f}%
r^{2}dr$. Since $\partial _{r}\left( \sqrt{-g}Q_{1}^{tr}\right) =0$, one can
match $\sqrt{-g}Q_{1}^{tr}$ at horizon and boundary. By calculating%
\begin{eqnarray}
\left. \sqrt{-g}Q_{1}^{tr}\right\vert _{\mathrm{horizon}} &=&-\frac{1}{2}ST+%
\frac{1}{2}\Lambda \Phi _{\Lambda },  \nonumber \\
\left. \sqrt{-g}Q_{1}^{tr}\right\vert _{\mathrm{boundary}} &=&-\frac{1}{4}M+%
\frac{1}{4}Q_{e}\Phi _{e}+\frac{1}{2}\Lambda \Phi _{\phi },
\end{eqnarray}%
we retrieve eq. (\ref{Smarr1}).

On the other hand, there is a global scaling symmetry for static gravity
theories \cite{Banados2005,Lu1507} which can be uncovered easily with a
different metric ansatz
\begin{equation}
ds^{2}=-u\left( \rho \right) dt^{2}+d\rho ^{2}+v\left( \rho \right) \left(
dx^{2}+dy^{2}\right) .
\end{equation}%
One can show that $L=\sqrt{-g}\left( L_{0}+L_{1}\right) $ with this metric
is invariant up to a total derivative $Q_{0}^{\prime }$, under the global
rescaling%
\begin{equation}
u\rightarrow \lambda ^{-2}u,\;v\rightarrow \lambda v,\;a_{t}\rightarrow
\lambda ^{-1}a_{t}.
\end{equation}%
Hence there is a Noether charge satisfied with $Q_{2}^{\prime }=0$:%
\begin{eqnarray}
Q_{2} &=&Q_{0}-\left( -2u\partial _{u^{\prime }}L+v\partial _{v^{\prime
}}L-a_{t}\partial _{a_{t}^{\prime }}L\right)  \nonumber \\
&&-2\left( -2u^{\prime }\partial _{u^{\prime \prime }}L+v^{\prime }\partial
_{v^{\prime \prime }}L-a_{t}^{\prime }\partial _{a_{t}^{\prime \prime
}}L\right) +\left( -2u\partial _{u^{\prime \prime }}L+v\partial _{v^{\prime
\prime }}L-a_{t}^{\prime }\partial _{a_{t}^{\prime \prime }}L\right)
^{\prime }.
\end{eqnarray}%
Note that the later two brackets are necessary since the Lagrangian depends
on higher derivative $u^{\prime \prime }$ and $v^{\prime \prime }$.
Substituting the concrete $L$ of massive gravity, we have%
\begin{equation}
Q_{2}=2\left[ -r^{2}\sqrt{fh}\left( \frac{2}{r}-\frac{h^{\prime }}{h}\right)
-Q_{e}a_{t}+c^{2}\tilde{\Phi}_{c}\right] ,
\end{equation}%
where $\tilde{\Phi}_{c}=\int 2\sqrt{h/f}dr$. Equating $Q_{2}$ at horizon and
boundary, which are%
\begin{eqnarray}
\left. Q_{2}\right\vert _{\mathrm{horizon}} &=&2TS-2c^{2}\Phi _{c},
\nonumber \\
\left. Q_{2}\right\vert _{\mathrm{boundary}} &=&3M-2Q_{e}\Phi _{e},
\end{eqnarray}%
leads to eq. (\ref{Smarr2}) again.

\section{Conclusion and discussion}

We derived the covariant form of the first law of black hole mechanics in
the presence of variable background fields. Due to its local nature, the
well-known expression of black hole entropy previously identified with a
Noether charge is still applicable, although the relevant diffeomorphism
symmetry is broken. The current situation is distinct from two important
works concerning the symmetry for the Noether entropy \cite%
{Tachikawa2006,Jacobson1507}. The former focused on the Chern-Simons term,
where the bare affine connection breaks not only the diffeomorphism symmetry
but also the general coordinate invariance, up to a total derivative. The
latter pointed out that in the frame formalism, both diffeomorphism and
Lorentz symmetries should be invoked. Combining these results, one may argue
that the diffeomorphism symmetry is not necessary nor sufficient to identify
what is\ the black hole entropy.

We illustrated the general formalism by a static black brane in the massive
EMD gravity. We derived the first law using the variational identity with
two variable background fields---the cosmological constant and the reference
metric. One can find that the conjugate variable of the cosmological
constant, the called black hole volume \cite%
{Kastor2009,Dolan1008,Mann2012,Mann2014,Liu2015,Cao2015}, involves the local
and nonlocal potentials. Such a separation suggests that these potentials
may have different physical origins by the gauge/gravity duality. We
identified two kinds of the scale invariance of the black brane solutions.
Using these together with the first law, we obtained two Smarr formulas. We
also proved both of them without invoking the first law.

Two implications from the present work deserve to be stressed. One is that
we provided a rare case for which the general coordinate invariance,
needless of the complete diffeomorphism symmetry, would induce nontrivial
physics: the general coordinate invariance has been used to greatly simplify
the covariant first law with background fields. We argue this is rare, since
the diffeomorphism symmetry that fathered the general relativity is always
taken seriously \cite{Misner}, but not the coordinate invariance alone which
any semi-respectable theory of physics can be required to respect \cite%
{Carroll}. In this regard, one may notice that the nonrelativistic version
of the general coordinate invariance developed in \cite{Son2006,Wu2015} has
been applied to construct the low-energy effective field theories of
condensed matter, such as the unitary Fermi gas and fractional quantum Hall
systems. Another example that is more alike to ours is the existence of the
possible conflicts between the dynamical and geometrical constraints for a
theory with explicit diffeomorphism breaking, which stems exactly from the
fact that the general coordinate invariance still holds \cite%
{Kosteleck2004,Bluhm2015}. Our work also suggests to regard the reference
metric and the cosmological constant in massive gravity as thermodynamic
variables, since the variation of each one corresponds to a Smarr formula
that has an explicit physical interpretation: one indicates the existence of
the generalized Killing potential and the other comes from a scaling symmetry
of the reduced action. We expect that the extension of the phase space in
massive gravity and the associated Smarr formulas would imply interesting results
in the black hole chemistry and holographic duals.

\begin{acknowledgments}
We are grateful to Hong Liu, Hong Lu and Sang-Jin Sin for valuable discussions. This
work was supported by NSFC (No. 11275120, 11375110, 11522541).
\end{acknowledgments}

\appendix

\section{Some tensors}

Here we list the explicit expressions of some tensors. The equations of
motion are%
\begin{eqnarray}
E_{\mu \nu }^{(g)}\!\! &=&\!\!\frac{\partial L}{\partial g^{\mu \nu }}\!-\!%
\frac{1}{2}g_{\mu \nu }L\!-\!X_{\;\;\;\;\;(\mu }^{\alpha \beta \rho }R_{\nu
)\rho \beta \alpha }  \nonumber \\
&&-2\nabla ^{\rho }\nabla ^{\lambda }X_{\lambda \left( \mu \nu \right) \rho
}\!+\!\nabla ^{\beta }A_{\left( \mu \nu \right) \beta }\!+\!\nabla ^{\beta
}B_{\left( \mu \nu \right) \beta },  \nonumber \\
E^{(\phi )}\!\! &=&\!\!\frac{\partial L}{\partial \phi }-\nabla _{\mu
}Y^{\mu },  \nonumber \\
E^{\left( a\right) \mu } &=&\frac{\partial L}{\partial a_{\mu }}-\nabla
_{\nu }Z^{\mu \nu },  \nonumber \\
E^{(b)\mu \nu }\!\! &=&\!\!\frac{\partial L}{\partial b_{\mu \nu }}-\nabla
_{\lambda }W^{\mu \nu \lambda },
\end{eqnarray}%
where the derivative of $L$ is involved, including%
\begin{eqnarray}
W^{\mu \nu \lambda }\!\! &=&\!\!\frac{\partial L}{\partial \nabla _{\lambda
}b_{\mu \nu }},~~~\;X^{\mu \nu \lambda \rho }=\frac{\partial L}{\partial
R_{\mu \nu \lambda \rho }},  \nonumber \\
Y^{\mu }\!\! &=&\!\!\frac{\partial L}{\partial \nabla _{\mu }\phi }%
,~~~\;\;\;\;\;\;\;\;Z^{\mu \nu }=\frac{\partial L}{\partial \nabla _{\nu
}a_{\mu }}.
\end{eqnarray}%
Two three-tensors are given by
\begin{eqnarray}
A_{\mu \nu \beta }\!\! &=&\!\!\frac{1}{2}\big(a_{\beta }Z_{\mu \nu }-a_{\mu
}Z_{\beta \nu }-a_{\nu }Z_{\mu \beta }\big), \\
B_{\mu \nu \beta }\!\! &=&\!\!\frac{1}{2}\big(W_{\alpha \mu \nu }b_{\;\beta
}^{\alpha }-W_{\alpha \beta \nu }b_{\;\mu }^{\alpha }-W_{\alpha \mu \beta
}b_{\;\nu }^{\alpha }  \nonumber \\
&&+W_{\mu \alpha \nu }b_{\beta }^{\;\alpha }-W_{\beta \alpha \nu }b_{\mu
}^{\;\alpha }-W_{\mu \alpha \beta }b_{\nu }^{\;\alpha }\big).  \label{Buvb}
\end{eqnarray}%
Note that we have arranged eq. (\ref{Buvb}) so that the first line and
second line in the parentheses are equal when $b_{\mu \nu }$ and $W^{\mu \nu
\lambda }$ are symmetric or antisymmetric on the $\mu \nu $ index.

The one-form $\theta =\theta \left( \psi ,\delta \psi \right) $ is given by%
\begin{eqnarray}
\theta ^{\beta }\!\! &=&\!\!2X_{(\mu \;\;\;\;\nu )}^{\;\;\;\alpha \beta
}\nabla _{\alpha }\delta g^{\mu \nu }-2\nabla _{\alpha }X_{(\mu \;\;\;\;\nu
)}^{\;\;\;\alpha \beta }\delta g^{\mu \nu }  \nonumber \\
&&+Y^{\beta }\delta \phi +Z^{\mu \beta }\delta a_{\mu }+W^{\mu \nu \beta
}\delta b_{\mu \nu }  \nonumber \\
&&-A_{\mu \nu }^{\;\;\;\;\beta }\delta g^{\mu \nu }-B_{\mu \nu
}^{\;\;\;\;\beta }\delta g^{\mu \nu }.
\end{eqnarray}%
The two-tensor $P_{\mu \nu }$ resulted from the general coordinate
invariance is
\begin{eqnarray}
P_{\mu \nu }\!\! &=&\!\!-2\frac{\partial L}{\partial g^{\mu \nu }}%
+4R_{\alpha \beta \rho \mu }X_{\ \ \ \;\;\nu }^{\alpha \beta \rho }+Y_{\nu
}\nabla _{\mu }\phi  \nonumber \\
&&+\left( Z_{\;\;\nu }^{\lambda }\nabla _{\mu }a_{\lambda }+Z_{\nu
}^{\;\lambda }\nabla _{\lambda }a_{\mu }+\frac{\partial L}{\partial
a_{\lambda }}a_{\mu }g_{\lambda \nu }\right)  \nonumber \\
&&+\big(W_{\;\;\;\;\nu }^{\alpha \beta }\nabla _{\mu }b_{\alpha \beta
}+W_{\nu }^{\;\alpha \beta }\nabla _{\beta }b_{\mu \alpha }+W_{\;\;\nu
}^{\alpha \;\;\beta }\nabla _{\beta }b_{\alpha \mu }\big)  \nonumber \\
&&+\frac{\partial L}{\partial b_{\lambda \rho }}\big(b_{\mu \rho }g_{\nu
\lambda }+b_{\lambda \mu }g_{\nu \rho }\big).  \label{puv}
\end{eqnarray}%
The three-tensor $\tilde{Q}_{\nu }^{\;\beta \alpha }$ in the Noether
potential $Q_{\xi }^{\beta \alpha }$ is%
\begin{eqnarray}
\tilde{Q}_{\nu }^{\;\beta \alpha }\!\! &=&\!\!\frac{1}{2}\Big(a_{\nu }Z^{%
\left[ \alpha \beta \right] }+a^{[\beta }Z_{\;\;\nu }^{\alpha ]}+a^{[\beta
}Z_{\nu }^{\;\alpha ]}  \nonumber \\
&&\!\!+W_{\nu \mu \;}^{\;\;[\alpha }b^{\beta ]\mu }+W_{\;\;\;\mu }^{[\alpha
\;\beta ]}b_{\nu }^{\;\mu }+W_{\;\;\;\mu \nu }^{[\alpha }b^{\beta ]\mu }
\nonumber \\
&&\!\!+W_{\;\;\nu }^{\mu \;[\alpha }b_{\mu }^{\;\beta ]}+W^{\mu \left[
\alpha \beta \right] }b_{\mu \nu }+W_{\;\;\;\;\;\nu }^{\mu \lbrack \alpha
\;\;}b_{\mu }^{\;\beta ]}\Big),
\end{eqnarray}%
where the last two lines are arranged similar to eq. (\ref{Buvb}).

\end{document}